\documentclass[aoas,preprint]{imsart}

\RequirePackage[OT1]{fontenc}
\RequirePackage{amsthm,amsmath}
\RequirePackage{natbib}
\RequirePackage[colorlinks,citecolor=blue,urlcolor=blue]{hyperref}
\usepackage{dsfont,resizegather,graphicx, multirow,enumitem, subcaption,float}

\newcommand\T{\rule{0pt}{2.6ex}}       
\newcommand\B{\rule[-1.2ex]{0pt}{0pt}}

\begin{document}

\begin{frontmatter}
\title{Predicting Melbourne Ambulance Demand using Kernel Warping}
\runtitle{Predicting Ambulance Demand using Kernel Warping}

\begin{aug}
\author{\fnms{Zhengyi} \snm{Zhou}\ead[label=e1]{zz254@cornell.edu}}
\and
\author{\fnms{David} \snm{Matteson}\ead[label=e2]{matteson@cornell.edu}}

\runauthor{Zhou and Matteson}

\affiliation{Cornell University}

\address{Center for Applied Mathematics\\
Cornell University\\
Ithaca, NY 14853\\
\printead{e1}\\
\phantom{E-mail:\ }}

\address{Department of Statistical Science\\
Cornell University\\
Ithaca, NY 14853\\
\printead{e2}\\
\phantom{E-mail:\ }}
\end{aug}

\begin{abstract}
Predicting ambulance demand accurately in fine resolutions in space and time is critical for ambulance fleet management and dynamic deployment. Typical challenges include data sparsity at high resolutions and the need to respect complex urban spatial domains. To provide spatial density predictions for ambulance demand in Melbourne, Australia as it varies over hourly intervals, we propose a predictive spatio-temporal kernel warping method. To predict for each hour, we build a kernel density estimator on a sparse set of the most similar data from relevant past time periods (labeled data), but warp these kernels to a larger set of past data irregardless of time periods (point cloud). The point cloud represents the spatial structure and geographical characteristics of Melbourne, including complex boundaries, road networks, and neighborhoods. Borrowing from manifold learning, kernel warping is performed through a graph Laplacian of the point cloud and can be interpreted as a regularization towards, and a prior imposed, for spatial features. Kernel bandwidth and degree of warping are efficiently estimated via cross-validation, and can be made time- and/or location-specific. Our proposed model gives significantly more accurate predictions compared to a current industry practice, an unwarped kernel density estimation, and a time-varying Gaussian mixture model.   
\end{abstract}



\end{frontmatter}

\section{Introduction} \label{sec:intro}

A primary goal of emergency medical services (EMS) is to minimize response times to life-threatening emergencies while keeping operational costs low. Accurate spatial-temporal ambulance demand predictions are crucial to optimal operations management of base location, staff, fleet, and deployment. These demand predictions are ideally needed at high temporal and spatial granularities. The industry typically predicts for every hour and every 1 km$^2$ region. We are motivated to predict this demand for the city of Melbourne, Australia.

There are several typical challenges to predicting ambulance demand. 
\begin{itemize}
\item Ambulance demand is often exceedingly sparse at the temporal and spatial resolution required for prediction. There is zero demand in the vast majority of 1-km$^2$ regions over a 1-hour period. 
\item This demand arises from complex urban geography. The city boundary is often highly irregular. Ambulance demand can be very high (coastal and downtown) or very low (suburbs) along the boundary. Within this boundary, demand follows closely the city's infrastructure and terrain; there might be high demand along central highways and zero demand within an internal lake. High resolutions covariates of these features are not readily available.
\item Ambulance demand exhibits spatial and temporal patterns. Weekly seasonality is usually prominent \citep{Channouf:2007, Matteson:2011}; the industry relies heavily on this seasonality to make predictions. Some studies have also noted daily seasonality and short-term serial dependence at densely-populated regions \citep{Zhou:2015a}.
\item Ambulance demand data for large cities is often large-scale. This presents computational challenges, especially since predictions are needed very frequently. 
\end{itemize}

It is particularly difficult to simultaneously resolve these challenges. Overcoming sparsity requires considerable smoothing, while capturing complex spatio-temporal patterns requires fine-resolution modeling. At high granularities, data sparsity makes it difficult to detect spatio-temporal characteristics accurately. At low granularities, differences across regions and times are not sufficiently captured for optimal ambulance planning. 

\begin{figure}[ht] 
\centering
\begin{subfigure}[a]{.46\textwidth}
 \centering
 \includegraphics[width=2.4in,height=1.6in]{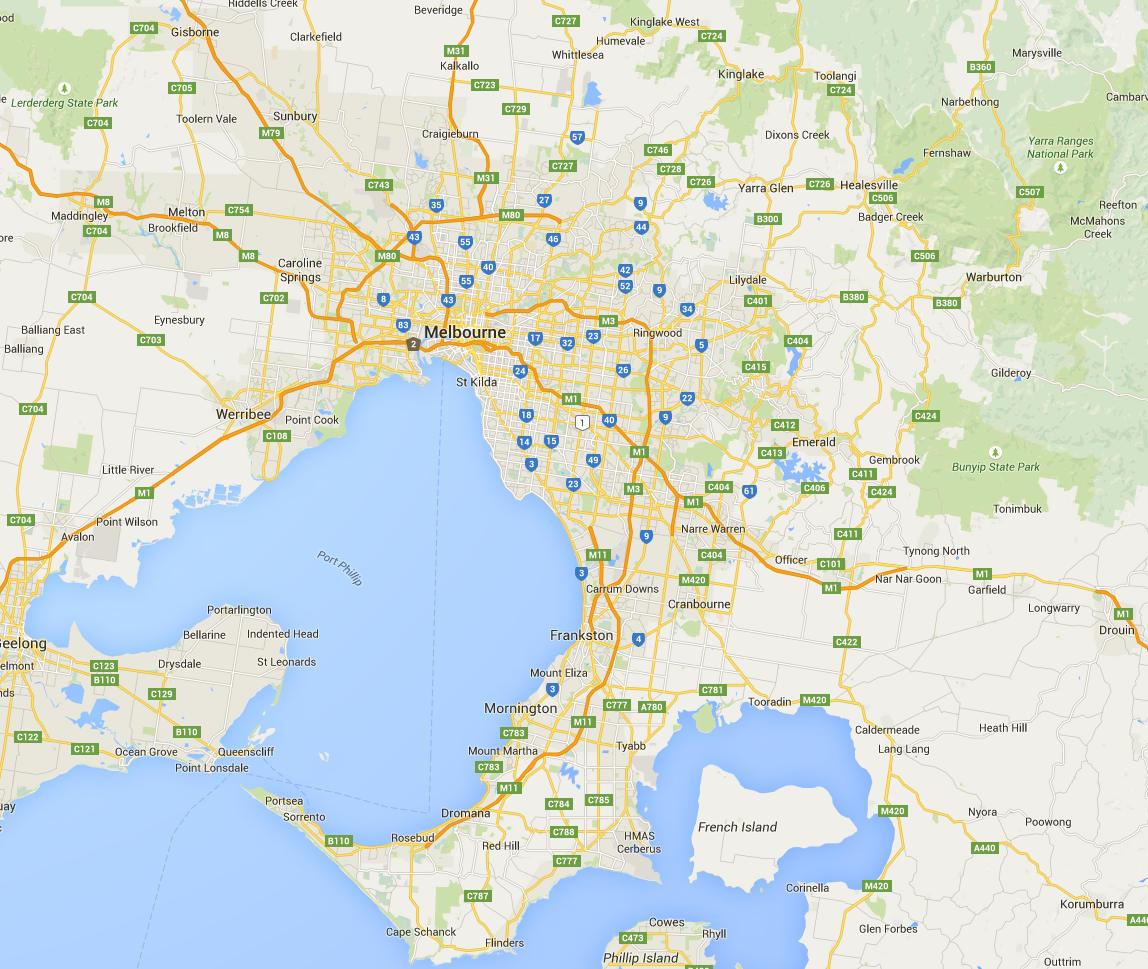}
\end{subfigure}
\quad
\begin{subfigure}[a]{.46\textwidth}
\centering
\includegraphics[width=2.7in,height=1.9in]{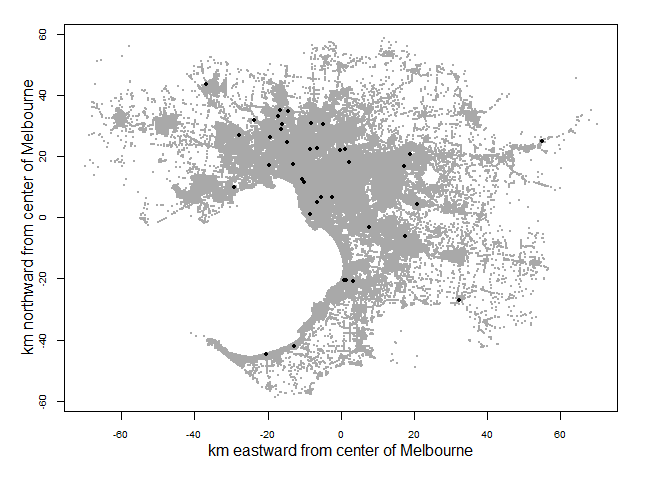}
\end{subfigure}
\caption{Left: map of Melbourne \citep{googlemap}; right: spatial locations of all $696,975$ Melbourne ambulance demand incidents from years 2011 - 2012 (in gray), and $38$ demand incidents for a typical 1-hour period (in black). We observe complex boundary and geographical features (e.g., highways, roads, satellite suburbs).}
\label{fig:data}
\end{figure}

Figure \ref{fig:data} demonstrates these challenges in predicting ambulance demand for Melbourne. We show on the right the locations of $696,975$ demand incidents from years 2011 and 2012 (in gray), and those of $38$ demand incidents for a typical 1-hour period (in black). On average, 99.6\% of the 1-km$^2$ regions in Melbourne receive zeros calls in any given hour. Comparing to a map of Melbourne on the left \citep{googlemap}, we observe a highly complex spatial boundary as Melbourne encloses a large bay to its south-west. Demand is high near the bay, but low on the outskirt suburbs. Demand is visibly higher at small satellite suburban neighborhoods and along major highways radiating out from the city center. There is lack of demand due to several reservoirs and a national park to the west and northwest. Consistent with typical patterns, the demand exhibits strong weekly seasonality.

The EMS industry and previous studies have attempted to address some of these challenges. The current industry practice uses a simple averaging formula. Demand in a 1-km$^2$ spatial region over an hour is typically predicted by averaging a small number of historical counts, from the same spatial region, over the corresponding hours from previous weeks or years \citep{Goldberg:2004}. For instance, the EMS of Charlotte-Mecklenburg, North Carolina uses a method called MEDIC, in which the prediction is the average of twenty corresponding counts in the same hour of the preceding four weeks for the past five years \citep{Setzler:2009}. Averaging so few historical counts, which are mostly zeros, produces noisy and flickering predictions, resulting in haphazard and inefficient deployment. 

Much attention has been given to predicting the aggregate ambulance demand as a temporal process, using autoregressive moving average models \citep{Channouf:2007}, factor models \citep{Matteson:2011}, and spectral analysis \citep{Vile:2012}. 
Few studies have modeled spatio-temporal ambulance demand well. \cite{Setzler:2009} use artificial neural networks, but fail to improve over the industry method. A recent study by \cite{Zhou:2015a} predicts ambulance demand for Toronto, Canada using a time-varying Gaussian mixture model (GMM). This method is more accurate than the industry practice, but, as the authors point out, extending it to incorporate spatial boundaries would be prohibitively expensive. While it may not be essential for Toronto since the city is almost rectangular in shape, it becomes important for Melbourne; it is difficult for ellipsoidal Gaussian components to model demand well on the highly complex spatial domain of Melbourne. Another study by \cite{Zhou:2015b} considers a spatio-temporal weighted kernel density estimation (KDE) to predict Toronto's ambulance demand. It gives similarly accurate predictions as GMM (both much better than industry), showing promise for KDE.

KDE has been used to analyze spatio-temporal data such as crime incidence \citep{Nakaya:2010}, disease spread \citep{Zhang:2011}, and data streams \citep{Aggarwal:2003}. It allows for rapid identification of ``hotspots'' and their evolutions in time and space. However, implementing a naive KDE is not satisfactory for our application. The chosen bandwidth necessarily has to be quite large given the data sparsity, smoothing inappropriately across boundary features and disregarding the underlying urban geography.

Few studies have focused on modeling spatial or spatio-temporal point processes on complex spatial structures. Most studies assume a boundary defined \textit{a priori} (polygon or pixelated). If not, \textit{ad hoc} methods based on the convex hull of all observed points are typically used \citep{Ripley:1977}.
This invariably results in a convex boundary that may be inaccurate where data is sparse. Even with a boundary optimally defined, few methods are equipped to handle complex boundary features. \cite{Ramsay:2002} proposes a finite window smoother with known boundary conditions computed using an expensive finite element approach. Building on that, \cite{Wood:2008} model the boundary condition as a loop of wire and the point process as a soap film suspended from the boundary wire. They represent this smoother as a penalized basis, compute it via multigrid, and select smoothness via generalized cross-validation. They acknowledge the lack of an elegant solution when the boundary conditions are unknown. Apart from boundary, other spatial characteristics, such as neighborhood structures and road networks, are rarely incorporated in modeling. We propose a method that can efficiently capture and exploit a wide range of spatial characteristics. We draw from theory and methods developed in manifold learning. 
 
Manifold learning, a branch of machine learning, is concerned with estimating and exploiting the underlying structures of data. The assumption is that data in a high-dimensional space resides on or near a lower-dimensional sub-manifold. In practice, we do not have access to this sub-manifold, but we can approximate it from a point cloud, i.e., a mass of historical data. The most common method is to construct an adjacency graph of this point cloud and make use of the properties and structures of this graph. This idea has led to many popular learning methods, including isomap \citep{Tenebaum:2000}, local linear embedding \citep{Roweis:2000}, and Laplacian eigenmaps \citep{Belkin:2003}. These methods were initially designed for data representation or visualization, but have been adapted for semi-supervised classification \citep{Belkin:2004}, and clustering \citep{Ng:2001, Shi:2000}. 

In particular, a variant of Laplacian eigenmaps, kernel warping, has been proposed for semi-supervised classification \citep{Smola:2003, Belkin:2004, Sindhwani:2005}. Using a small number of labeled data and a larger number of labeled and unlabeled point cloud data, the method classifies new examples by constructing kernels on the labeled data that warp to the geometry of the point cloud. This geometry is represented by the adjacency graph of the point cloud. Smoothing orthogonal to this geometry is penalized heavily, whereas smoothing along this geometry is not. This method is designed for high-dimensional classification, and has good performance on text and image data.

Drawing from this idea, we propose a novel method for modeling spatio-temporal point processes against complex spatial structures and features. To predict ambulance demand for a future time period, we have a sparse set of historical data that is very relevant for this prediction (labeled data). We fit a KDE on them, but warp the kernels to a larger set of historical data regardless of their relevance to this predictive task (point cloud). This point cloud describes our belief about the spatial structure on which the labeled data lies. It captures exterior and interior boundaries without needing to explicitly define boundaries and boundary conditions. It also incorporates a wide range of complex spatial similarities and discontinuities, such as roads, city blocks, and neighborhoods of varying shapes and densities. Intuitively, this warping can be thought of as a regularization that penalizes radical departure from and encourages flow of information along our intuition of the geography. In a Bayesian sense, it can also be thought of as imposing a prior based on how similar or different the point process is across different locations. Such a regularization or prior is especially beneficial when the labeled data is sparse. We select the kernel bandwidth and the degree of warping efficiently via cross-validation. Both of these parameters can be made time- and/or location-specific. 

We implement this method on ambulance demand data from Melbourne in years 2011 and 2012. Altogether there are $696,975$ realized events. Each event contains the time and location that the ambulance was dispatched to. The proposed kernel warping model gives significantly more accurate predictions than previous approaches, including the MEDIC method as an industry practice, unwarped KDE, and GMM. 

We develop the kernel warping model in Section \ref{sec:model}. We construct an unwarped KDE in \ref{sec:kde}, warp the kernels to the point cloud in \ref{sec:warp}, and allow for time- and location-specific warping in \ref{sec:stwarp} for the Melbourne data. Some details on computation are included in \ref{sec:comp}. We show the empirical results for predicting Melbourne ambulance demand in Section \ref{sec:result}, and conclude in Section \ref{sec:concl}. 

\section{Model} \label{sec:model}
We model Melbourne's ambulance demand on a continuous spatial domain $\mathcal{S}\subseteq \mathds{R}^2$  and a discretized temporal domain of one-hour intervals $\mathcal{T}=\{1,2, \ldots\}$. Let $\mathbf{s}_{t,i}$ be the location of the $i$-th ambulance demand arising from the $t$-th time period, for $i\in \{1,\ldots,n_t\}$, where $n_t$ is the total number of ambulances demanded in the $t$-th period. Since a non-homogeneous Poisson process (NHPP) is a natural model for spatial point process \citep{Diggle:2003, Moller:2004}, we assume $\{\mathbf{s}_{t,i} : i = 1,\ldots n_t\}$ for each time period $t$ independently follow an NHPP over $\mathcal{S},$ with positive intensity function $\lambda_t$. 
We decompose the intensity function as $\lambda_t(\mathbf{s})=\delta_t f_t(\mathbf{s})$, for $\mathbf{s}\in\mathcal{S}$. Here, $\delta_t=\int_{\mathcal{S}} \lambda_t(\mathbf{s})\, d\mathbf{s}$ is the aggregate demand intensity over the spatial domain, and $f_t(\cdot)$ is the continuous spatial \textit{density} of the demand at time $t$ such that $f_t(\mathbf{s})>0$ and $\int_\mathcal{S} f_t(\mathbf{s})d\mathbf{s}=1$. Therefore, for each $t$, $n_t | \lambda_t \sim \mbox{Poisson}(\delta_t)$ and $\mathbf{s}_{t,i} | \lambda_t, n_t \stackrel{iid}{\sim} f_t(\cdot)$ for $i \! \in \! \{1,\ldots,n_t\}$. The usual practice is to model $\{\delta_t\}$ and $\{f_t\}$ separately. As mentioned before, numerous studies have proposed sophisticated and accurate methods for estimating $\{\delta_t\}$. We thus focus on predicting the spatio-temporal demand density $\{f_t\}$, which is more challenging and less studied. 

\subsection{Spatio-temporal KDE} \label{sec:kde}

Suppose we want to predict Melbourne's ambulance demand for a future 1-hour period $u$. Given the prominent weekly seasonality, the most relevant observations are from the corresponding hour of the week for the past $M$ weeks. They constitute the labeled data for this predictive task. This approach is aligned with the industry practice, and is shown to work well in \cite{Zhou:2015a}. We choose the sliding window width $M$ \textit{a priori}. With a larger $M$, we have more training data, but each training is slower and less adaptive to recent changes in demand patterns (e.g., summer vs. winter). The industry and recent studies have considered $M$ between 4 and 8. For Melbourne, we set $M=8$, resulting in an average labeled data size of about 300 points (ranging from 100 to 450 for different periods). Let $\mathcal{T}_{u}=\{u-168m: m\in\{1,\ldots,M\}\}$ denote the set of labeled time periods, in which 168 is the number of 1-hour periods in a week.

Starting with a simple KDE on the labeled data, we predict for any $\mathbf{x}\in \mathcal{S}$,
\begin{equation}\label{eqn:kde}
f_{u}(\mathbf{x}) =\frac{1}{\sum_{t \in \mathcal{T}_{u}}n_t}\sum_{t \in \mathcal{T}_{u}} \sum_{i=1}^{n_t} k\!\left(\mathbf{x}, \,\mathbf{s}_{t,i} \,\vert\, \boldsymbol{H}\!\right).
\end{equation}
Here, $k$ is the chosen bivariate spatial kernel with bandwidth matrix $\boldsymbol{H}$.
We use the Gaussian kernel, and choose bandwidth $\boldsymbol{H}$ via the plug-in method \citep{Wand:1994} or smoothed cross-validation \citep{Duong:2005}. When data show large variations in density, using one fixed bandwidth may not be optimal \citep{Cacoullos:1966, Scott:1992}. A bandwidth too large wipes out local features where we have sufficient data; a bandwidth too small leads to spurious peaks where data is sparse. In the case of Melbourne, data density varies substantially in space (downtown vs. neighborhoods) and time (midnight vs. rush hours); we may be motivated to consider a spatial- and/or time-varying $\boldsymbol{H}$. 

\subsection{Kernel warping} \label{sec:warp}
We would like to warp each kernel $k$ in Equation (\ref{eqn:kde}) to a larger set of point cloud data that describes the spatial boundary and characteristics of Melbourne. We choose the point cloud data, construct an adjacency graph on the point cloud, define the graph Laplacian matrix, and warp the kernel to this Laplacian matrix. We discuss in detail each step.

\noindent \textbf{Step 1 [Choosing the point cloud]}: Typically in Laplacian eigenmap and kernel warping applications, all labeled and unlabeled data is used as the point cloud. In the context of spatial statistics and our application, there are several points of consideration:
\begin{enumerate}[label=(\alph*)]
\item \textit{Which points?} We consider all observations in the near past, regardless of the time period. If we use the same sliding window width of $M = 8$ previous weeks, we are choosing from about 50,000 points.
\item \textit{How many points?} There is a trade-off: using more points in the cloud leads to better approximation of the geography but slower computation. Since we are in a low-dimensional space of $\mathds{R}^2$, we may not need a very large number of points to depict the most salient boundary and spatial structures. In our application, we find 1000 spatial points represent Melbourne's geography reasonably well. 
\item \textit{Points or mesh?} Alternative to using past observations, we can also use past data to define a pixelated spatial domain of Melbourne and use the included pixels as the point cloud. Doing so we lose some resolution and information on data density, but may gain computationally if it can reduce the number of point cloud data significantly. A regularly spaced point cloud also induces a sparse, band-diagonal graph Laplacian matrix (to be discussed later), leading to further savings. 
\item \textit{Global or local?} We can have one global point cloud for the entire spatial domain. We can also discretize the spatial domain into several regions with separate local point clouds. Local point clouds can provide computational advantages if they are smaller. They may also offer accuracy advantages if they depict finer-grain characteristics or allow for customized degree of warping at each locale. We discuss this further in Section \ref{sec:stwarp}.
\end{enumerate}

\begin{figure}[ht] 
\centering
\includegraphics[width=1\linewidth]{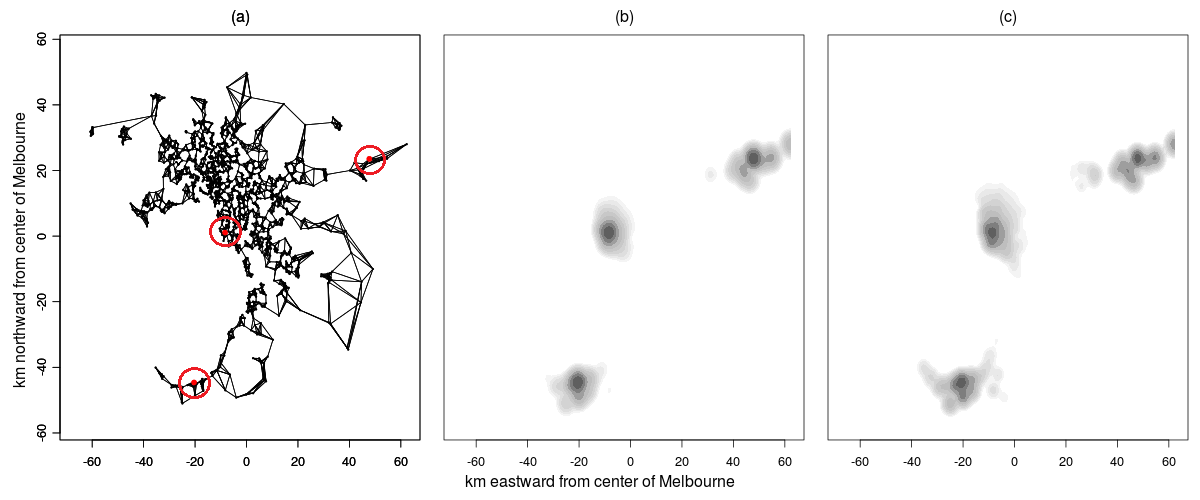}
\caption{Examples of kernel warping: (a) the adjacency graph of a sample point cloud of size 1000; three observations are highlighted in red; (b) and (c), warped kernels centered at the these three observations with degrees of deformation $\lambda = 0.5$ and $2$, respectively. }
\label{fig:exp}
\end{figure}

In our application, we randomly sample 1000 historical observations as the point cloud for each ``component'' (to be explained in Section \ref{sec:stwarp}). We denote the set of point cloud data as $\{\boldsymbol{z_i}\}$ for $i\in \{1,\ldots,Z\}$. See Figure \ref{fig:exp} (a) for an example cloud of 1000 points over the entire city of Melbourne. For our application, we find that predictive accuracy is not sensitive to the random sampling of the point cloud data. If it were, a larger point cloud might be needed, or predictions might be repeated and averaged over several point cloud samples. \\ 

\noindent \textbf{Step 2 [Constructing the adjacency graph]}: We construct a graph with nodes at each point in the point cloud and edges connecting points that are close. We represent this graph using a symmetric, positive semidefinite adjacency matrix $A$. 
\begin{enumerate}[label=(\alph*)]
\item \textit{Which nodes to connect?} Knowledge about the spatial domain (e.g., inside a building vs outside) or regularity of the point cloud (e.g., regular mesh) may inform a natural way to define how nodes should be connected. Without such knowledge, we can connect nodes $\boldsymbol{z}_i$ and $\boldsymbol{z}_j$ if $\boldsymbol{z}_i$ is among the $n$ nearest neighbors of $\boldsymbol{z}_j$ \textit{or} $\boldsymbol{z}_j$ is among the $n$ nearest neighbors of $\boldsymbol{z}_i$ (symmetric relation). This requires us to choose $n$. In our experience, $n$ should be big enough to ensure that the point cloud is sufficiently connected instead of being very fragmented, but small enough to emphasize local relationships. A second way is to connect nodes if the (Euclidean) distance between them is smaller than a threshold. This requires us to choose the threshold. 
\item \textit{Weighted edges?} In the simplest case, we can set $A_{ij}=1$ if nodes $\boldsymbol{z}_i$ and $\boldsymbol{z}_j$ are connected and 0 otherwise. Another idea suggested in \cite{Belkin:2003} is to define weighted edges depending on the distance between points, i.e., $A_{ij}= \exp\{-||\boldsymbol{z}_i-\boldsymbol{z}_j||^2 / r\}$ if $\boldsymbol{z}_i$ and $\boldsymbol{z}_j$ are connected and 0 otherwise. The authors note that they do not have a principled way of choosing $r$; we find it reasonable to choose $r$ empirically by fitting an exponential distribution on all distances between connected nodes. They also note that in practice a binary adjacency graph works well, and we agree. 
\end{enumerate}

In the Melbourne application, we use $n=5$ nearest neighbors and binary weights to construct $A$. Figure \ref{fig:exp} (a) shows the adjacency graph of a sample point cloud of size 1000. Again, we find our predictive accuracy to be insensitive towards any reasonable variations in these choices.\\

\noindent \textbf{Step 3 [Constructing the Laplacian matrix]}: The graph Laplacian matrix $L$ is defined to be $L=D-A$, in which $D$ is the diagonal degree matrix, with its diagonal entries being the column (or equivalently, row) sum of $A$, i.e., $D_{ii}=\sum_j A_{ij}$. $L$ is a symmetric, positive semidefinite matrix. If the graph has multiple connected components, $L$ can be rearranged into a block diagonal matrix, where each block is the respective Laplacian matrix for each connected component. 

Here is the intuition of the Laplacian matrix. The (discrete) point cloud adjacency graph is an empirical approximation to our target (continuous) manifold of Melbourne geography. The (discrete) graph Laplacian matrix $L$ is then an approximation to the (continuous) Laplace-Beltrami operator on this manifold. The Laplace-Beltrami operator is a manifold generalization of the Laplace operator, which is a linear second order differential operator on functions (in our case, kernels). This $L$ induces a semi-norm on kernels which penalizes changes between adjacent nodes. There is a close analogy to heat flow; the heat (partial differential) equation has a Laplace operator in space. Intuitively, $L$ guides how information (heat) spreads on the spatial structure (manifold approximated by graph) from any initial KDE (initial heat distribution). \\

\noindent \textbf{Step 4 [Warping the kernels]}: We warp each kernel $k$ from Equation (\ref{eqn:kde}) to the point cloud to generate a new warped kernel $\tilde k$. For any $\mathbf{x}\in\mathcal{S}$ and any $\mathbf{s}$ in the set of labeled data
\begin{equation}\label{eqn:warp}
\tilde k(\mathbf{x},\mathbf{s} \,\vert\, \boldsymbol{H})=k(\mathbf{x},\mathbf{s}\,\vert \, \boldsymbol{H})-\boldsymbol{k}_\mathbf{x}^T (\boldsymbol{I}+\lambda L\boldsymbol{K})^{-1} \lambda L\boldsymbol{k}_\mathbf{s},
\end{equation}
in which $\boldsymbol{k}_\mathbf{x}= \left[k(\mathbf{x},\mathbf{z}_1 \,\vert \, \boldsymbol{H}),\ldots,k(\mathbf{x},\mathbf{z}_Z \,\vert \, \boldsymbol{H})\right]$ and $\boldsymbol{k}_\mathbf{s}=\left[k(\mathbf{s},\mathbf{z}_1 \,\vert \, \boldsymbol{H}),\ldots,\right.$ \linebreak $\left.k(\mathbf{s},\mathbf{z}_Z\,\vert \, \boldsymbol{H})\right]$ are vectors of kernels evaluated at $\mathbf{x}$ or $\mathbf{s}$ and the point cloud data $\{\boldsymbol{z}_i\}$. Matrix $\boldsymbol{K}=\left[k(\mathbf{z}_i, \mathbf{z}_j \,\vert \, \boldsymbol{H})\right]_{i,j\in\{1,\ldots,Z\}}$ is a symmetric matrix of kernels evaluated at all pairs of point cloud data, and $\boldsymbol{I}$ is a $Z$ by $Z$ identity matrix. The parameter $\lambda>0$ represents the degree of deformation. When $\lambda=0$, we have $\tilde k=k$. When $\lambda \rightarrow \infty$, $\tilde k$ approaches a positive constant on the point cloud (steady state heat distribution).

Equation (\ref{eqn:warp}) is obtained by warping the Reproducing Kernel Hilbert Space (RKHS) associated with the chosen kernel. We modify the RKHS with a point-cloud semi-norm $\lambda L$. This deforms the kernel $k$ along a finite-dimensional subspace given by the point cloud data. The modified RKHS is shown to be another RKHS, i.e., $\tilde k$ is a properly defined kernel. See \cite{Sindhwani:2005} and \cite{Belkin:2006} for more details (they use the point cloud semi-norm of $\lambda L^p$; we consider the simplified case where $p=1$). 

There are three interpretations of this type of kernel warping. The first is that of heat flow as mentioned before. We allow information (heat) to spread along the graph of the point cloud (approximately the manifold of Melbourne's geography).  The second interpretation is a graph regularizer. Variations between adjacent nodes in the graph are penalized, and thus violation of the spatial structure implied by the point cloud are penalized. Lastly, in the Bayesian framework, kernel warping can informally be thought of as imposing a data-dependent informative prior to describe our belief of the data geometry.

We replace the regular Gaussian kernel $k$ in Equation (\ref{eqn:kde}) with the new warped kernel $\tilde k$ defined in (\ref{eqn:warp}) to predict the density of ambulance demand $f_u$ at a future time period $u$. We set \textit{a priori} the sliding window width $M$, the point cloud data type / size, the number of nearest neighbors $n$, and the weights used to construct the Laplacian matrix. We estimate the Gaussian kernel bandwidth $\boldsymbol{H}$ and the degree of deformation $\lambda$. 

We show in Figure \ref{fig:exp} (b) and (c) examples of warping kernels. Three kernels of bandwidth $\boldsymbol{H}=\mbox{diag}(2,2)$ are placed on three observations circled in red in Figure \ref{fig:exp} (a). They are warped towards the point cloud in (a) with degree of deformation $\lambda =0.5$ (b) and $2$ (c). With a larger $\lambda$, the kernels conform to the spatial boundary and characteristics to a greater extent. 

\subsection{Spatio-temporal kernel warping} \label{sec:stwarp}
Melbourne's ambulance demand shows substantial density variations with patterns in time (midnight vs rush hour) and in space (downtown vs neighborhoods). It may be beneficial to allow bandwidth $\boldsymbol{H}$ and degree of deformation $\lambda$ to vary with time and space. Ideally, we would like to find, in time and space, pockets of the point process with similar characteristics, and apply similar smoothing and deformation. 

We discretize time according to our modeling aims, i.e., into 1-hour time periods. For each hour, we further discretize the spatial domain into a small number of regions, as motivated by the behavior of labeled data for that time period. We call each subregion of each hour a \textit{component}, and perform estimations and predictions independently on each component. The spatial discretization splits a global point cloud into local ones, cuts all edges connecting across regions, and decomposes the Laplacian matrix into blocks. Labeled data are also matched into components. We estimate a separate set of $\boldsymbol{H}$ and $\lambda$ for each component by cross-validation (details in Section \ref{sec:comp}).

We discretize spatially by clustering. For any given future time period, we cluster on its labeled data (about 300 observations). We allow different numbers of clusters and clustering configurations for each time period. In our application, this gives more accurate predictions than imposing a universal clustering configuration across time. We also obtain better results by clustering on labeled data rather than clustering on the point cloud data (the point cloud is much more similar across time than the labeled data). In the case of Melbourne, spatial characteristics across time are different enough that the gain in personalized modeling exceeds the loss in stablization offered by a common arrangement. 

We choose to cluster using K-means based on Euclidean distance. K-means is fast, clusters all points, and gives even clusters. Even cluster sizes are desirable because a very small cluster does not provide enough labeled data to reliably estimate parameters via cross-validation. To avoid this, we set a threshold minimum number of points in any cluster. We set the threshold at 15 points, which in practice limits the number of clusters to be below 8. If we fail to clear this threshold, we lower the number of clusters. Density-based clustering algorithms such as DBSCAN \citep{Ester:1996} and shared nearest neighbors \citep{Ertoz:2003} do not classify all points, and do not allow easy specification of the number of clusters. Graph-based clustering algorithms such as affinity propagation \citep{Frey:2007} and spectral clustering \citep{Ng:2001} do not cluster on Euclidean distance, and may be less intuitive for spatial point patterns. In our case, hierarchical clustering gives very uneven cluster sizes. 

For each time period, we binary search for the best number of clusters based on validation likelihood. Increasing the number of clusters leads to, on the one hand, an additional 1000 points to the cloud and the flexibility to customize parameters locally, but on the other hand, sparser labeled data for each cluster and reduced stability in parameter estimation. It is an empirical question for each time period whether we have enough labeled data to afford this increase in complexity. For Melbourne, we find the number of clusters to be largely proportional to the size of labeled data.

\subsection{Computation} \label{sec:comp}

We estimate the kernel bandwidth $\boldsymbol{H}$ and the degree of deformation $\lambda$ for each spatio-temporal component. To reduce the dimensionality, we parametrize $\boldsymbol{H}$ to be a scalar multiple of the plug-in bandwidth $\boldsymbol{H}_{pi}$ obtained if we fit an unwarped KDE for the same component. That is we define $\boldsymbol{H}=\alpha \boldsymbol{H}_{pi}$, and estimate $\alpha$. Alternatively, we can define a radial bandwidth $\boldsymbol{H}=\mbox{diag} (\beta,\beta)$, reducing the Gaussian kernel to a radial basis function. We use $\boldsymbol{H}=\alpha \boldsymbol{H}_{pi}$ because this parametrization gives slightly better performance in our preliminary analysis. To estimate a full $\boldsymbol{H}$ is more difficult because $\boldsymbol{H}$ needs to be positive semi-definite. 

We choose $\boldsymbol{H}$ and $\lambda$ for each component using 5-fold cross-validation to maximize average validation likelihood. We implement a surrogate, derivative-free optimization procedure called the stochastic radial basis function (RBF) method \citep{Regis:2007, Regis:2009}. It is a fast algorithm for global optimization of computationally expensive objective functions. Each iteration builds an RBF model to approximate the expensive function, selects subsequent candidate points, and evaluates them in parallel. We choose this approach because our objective function (likelihood) evaluation is not instantaneous. It takes between 0.5 and 4 seconds, depending on the sizes of the labeled data and point cloud (Python code on a personal computer). We also do not have simple derivative computations. In our experience, 100 such evaluations are sufficient to provide a good optimum, competitive to those found by grid search, pattern search, or evolutionary algorithms. However, a wide range of optimization tools can be applied here. 

In our application, we find a typical optimal $\alpha$ to be between 0.05 and 0.3. We need a concentration of heat which is then spread or warped to the point cloud. A typical optimal $\lambda$ is between 0 and 2. Most time periods choose between 1 and 3 spatial components. We warm start the binary search for the number of clusters based on the size of labeled data. The best configuration is usually found within 3 searches. 

Given the prominent weekly seasonality, we believe that the corresponding parameter values are also similar from week to week. In fact, we believe that the nature of deformation and smoothing does not vary significantly over several months, and thus only estimate the parameters for a one-week cycle once every few months. With the most recent weekly set of parameter values, we predict forward in an online fashion with a sliding window of $M=8$ weeks, making use of the most recent 8 weeks of data available. Each prediction is instantaneous. 

The most expensive part of the computation is evaluating kernels between all pairs of point cloud data and taking the inverse of a large matrix. Several local point clouds of reasonable sizes ($<$ 2000) is computationally more efficient than one massive global point cloud. There are ways to optimize this computation, including using right division instead of inversion, saving pre-computed kernel evaluation matrices and vectors, exploiting sparse, banded-diagonal Laplacian matrix, using a tree-based algorithm for fast KDE computation \citep{Gray:2003}, and using a look-up table for Gaussian densities (most of these optimizations are not used in our implementation). The computation is ``embarrassingly parallelizable'', across validation likelihood evaluations and across spatio-temporal components.

\section{Predicting ambulance demand for Melbourne} \label{sec:result}

We would like to predict ambulance demand in Melbourne for every 1-hour period in March 2011. There are two stages to this computation. In the first stage, we estimate all parameters for a weekly cycle. The parameters include the spatial clustering configuration for each 1-hour period, as well as the parameters $\lambda$ (degree of warping) and $\alpha$ (in bandwidth $\boldsymbol{H}=\alpha \boldsymbol{H}_{pi}$) for each spatial component in each 1-hour period. This estimation only needs to be performed very infrequently, and in our case, once. For this estimation, we use Melbourne ambulance demand data from 8 weeks in January and February 2011. In the second stage, we use the estimated weekly set of parameter values to predict future ambulance demand on a sliding window of 8 weeks for each 1-hour period in March 2011. 

\begin{figure}[ht] 
\centering
\includegraphics[width=\linewidth]{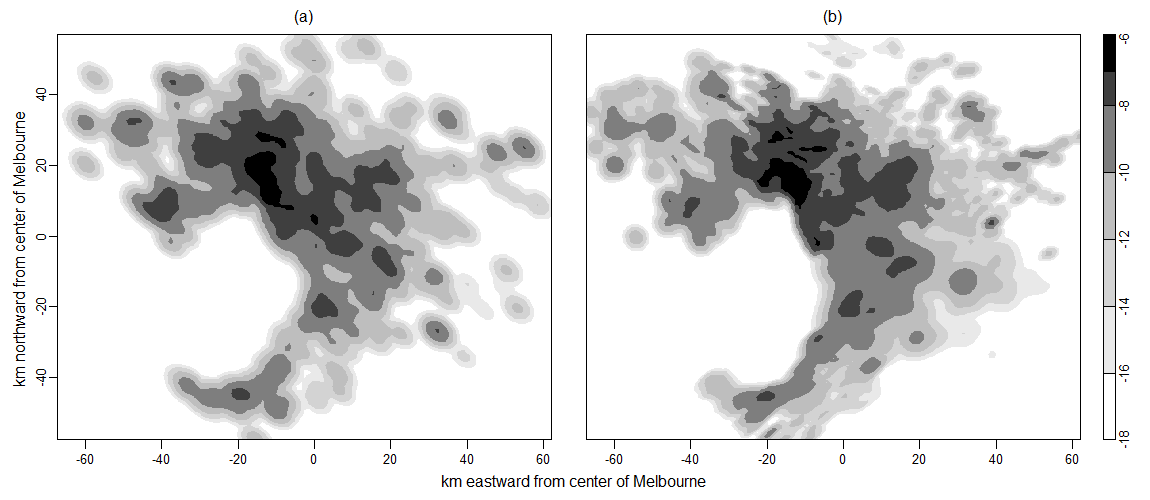}
\caption{Log predictive densities using spatio-temporal kernel warping for March 2, 2011 (Wednesday) at (a) 2 - 3 am (night), and (b) 2 - 3 pm (day). For time period (a), we have sparse data and cross-validate to choose 1 spatial component. For time period (b), we have more data and choose 5 spatial components.}
\label{fig:predden}
\end{figure}

Figure \ref{fig:predden} shows the predictive density estimated by kernel warping for two time periods on March 2, 2011 (Wednesday). We have only about 150 labeled data to predict for 2 - 3 am (a), and cross-validate to use only 1 spatial component. We have almost 400 labeled data for 2 - 3 pm (b) and cross-validate to choose 5 spatial components.  

\begin{figure}[ht] 
\centering
\includegraphics[width=\linewidth]{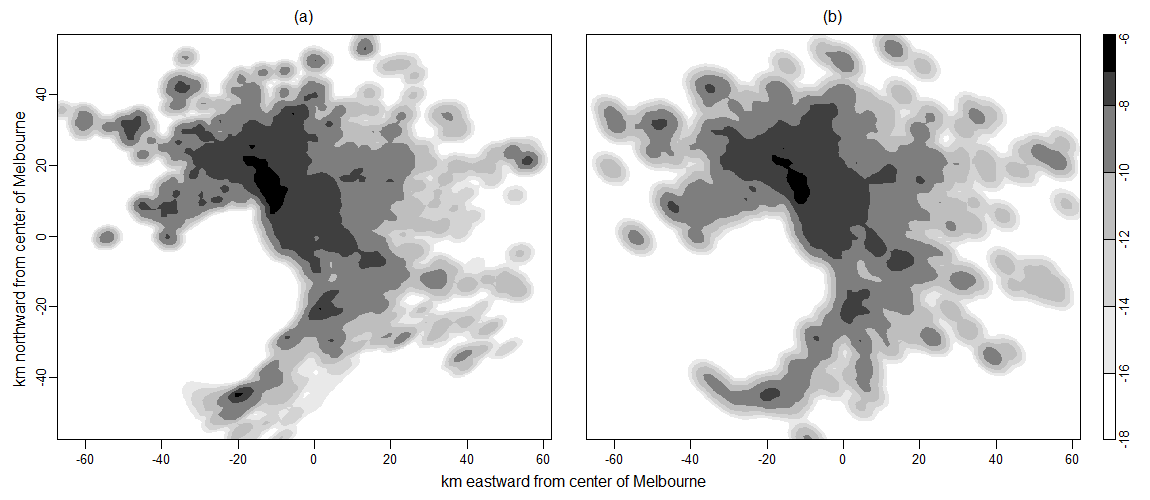}
\caption{Log predictive densities for March 2, 2011 (Wednesday) at  10 - 11 am  using (a) spatio-temporal kernel warping (3 spatial clusters), and (b) temporal kernel warping. The density in (a) shows slightly more details.}
\label{fig:stvst}
\end{figure}

We consider two variations in estimation: (i) spatio-temporal kernel warping (S-T param), in which we separately estimate parameters for each 1-hour period and spatial region (via clustering, Section \ref{sec:stwarp}), and (ii) temporal kernel warping (T param), in which we separately estimate parameters for each 1-hour period (no spatial clustering). We show in Figure \ref{fig:stvst} the predictive densities produced by these two approaches for the same time period. The densities look similar, with slightly more details when we use spatio-temporal kernel warping (we cross-validate to select 3 spatial clusters).

We compare the proposed kernel warping models to the following 
\begin{enumerate}[label=(\alph*)]
 \item The MEDIC method, which is an industry practice implemented in Charlotte-Mecklenburg, NC (Section \ref{sec:intro}). We implement this method as far as we have data. The cell count in a 1-km$^2$ region and a 1-hour period is predicted by the average of corresponding cell counts in the preceding 8 weeks.  
 \item Unwarped KDE, as in Equation (\ref{eqn:kde}). The bandwidth $\boldsymbol{H}$ is chosen via the plug-in method (PI) \citep{Wand:1994} and smoothed cross-validation (SCV) \citep{Duong:2005}. This $\boldsymbol{H}$ is separately estimated for each time period, but does not vary in space.
 \item Gaussian mixture model (GMM) \citep{Zhou:2015a}, in which the means and covariances of Gaussian components are fixed through time, and the mixture weights vary in time. We also use labeled data from the last 8 weeks, and consider 15, 30 and 50 components. The computational expenses are substantial. 
\end{enumerate}

\begin{figure}[ht] 
\centering
\includegraphics[width=\linewidth]{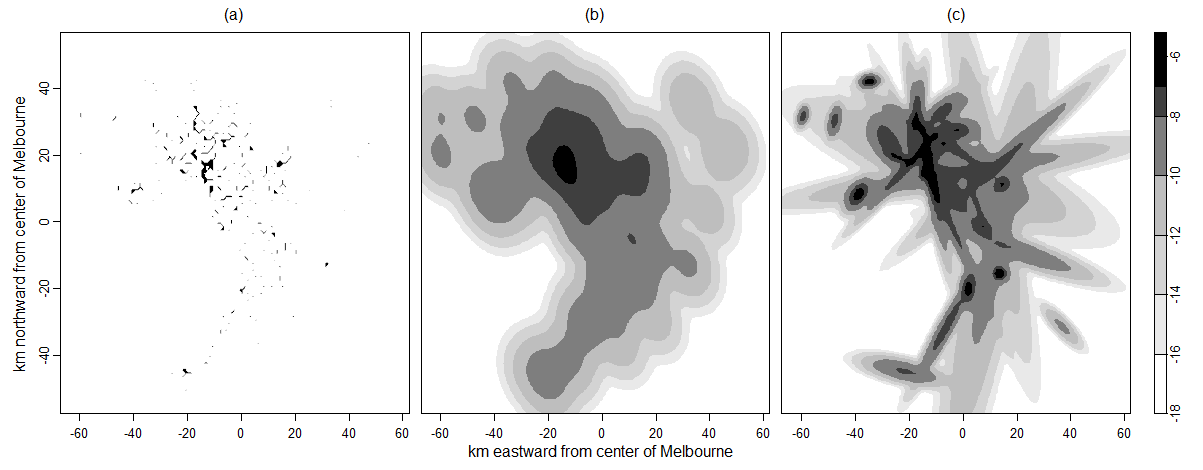}
\caption{Log predictive densities using comparison methods for 2 - 3 pm on March 2, 2011 (Wednesday): (a) the MEDIC method (an industry practice); (b) unwarped KDE with bandwidth selected by the plug-in method (PI); (c) time-varying Gaussian mixture model with 30 components. These densities are to be compared to Figure \ref{fig:predden} (b), which is the prediction using kernel warping for the same period.}
\label{fig:comparison}
\end{figure}

Figure \ref{fig:comparison} shows the log predictive density using the MEDIC method, unwarped KDE (PI), and GMM (30 components) for March 2, 2011 at 2 - 3 pm. These densities are comparable with Figure \ref{fig:predden} (b), which shows the log predictive density for the same period predicted by the proposed kernel warping. Even with 400 labeled data, The MEDIC method gives exceedingly noisy predictions, while unwarped KDE and GMM produce over-smoothed densities that do not adapt well to the spatial features of Melbourne.

We use several performance metrics to compare the statistical predictive accuracies of different methods. First, we use average log score (ALS) \citep{Good:1952}. This metric is advocated for being a strictly proper scoring rule closely related to Bayes factor and Bayes information criterion \citep{Gneiting:2007}. It is the average log likelihood of test data. For each test time period $u$ in the set of all test time periods $\mathcal{T}_{test}$,
\[
\mbox{ALS } (u)=\sum_{i=1}^{n_u}\log \hat f_u(\tilde{\mathbf{s}}_{u,i}),
\]
in which $\{\tilde{\mathbf{s}}_{u,i}\}$ are the test data, and $\hat f_u(\cdot)$ is the predictive density for period $u$ obtained by various methods. For the MEDIC method, we normalize cell counts to discrete density by dividing over the total count in each period. 

Secondly, we compare accuracy in cell counts for every 1-km$^2$ region and 1-hour period. For the proposed kernel warping, unwarped KDE, and GMM, we discretize continuous predictions in space to each 1 km$^2$, and convert to counts by multiplying the total count for the period as predicted by the MEDIC method. We compute the root-mean-square error, both within the smallest rectangle enclosing all data (plotting window in Figures \ref{fig:data}, \ref{fig:predden} - \ref{fig:comparison}) (RMSE) and within a pixelated data-driven boundary of Melbourne $B$ ($\mbox{RMSE}_B$). For each test time period $u\in \mathcal{T}_{test}$,
\[
 \mbox{RMSE }(u)=\sqrt{\frac{1}{C}\sum_{c=1}^{C}(y_{u,c}-\hat y_{u,c})^2},
\]
where $C$ is the number of 1 km$^2$ cells in the rectangular observation window, $y_{u,c}$ and $\hat y_{u,c}$ are the actual and predicted count for period $u$ and cell $c$ respectively. For $\mbox{RMSE}_B$, we use cells $c$ within the pixelated boundary $B$ and $C$ as the number of 1 km$^2$ cells within this boundary.

\begin{figure}[ht] 
\centering
\includegraphics[width=\linewidth, height=1.8in]{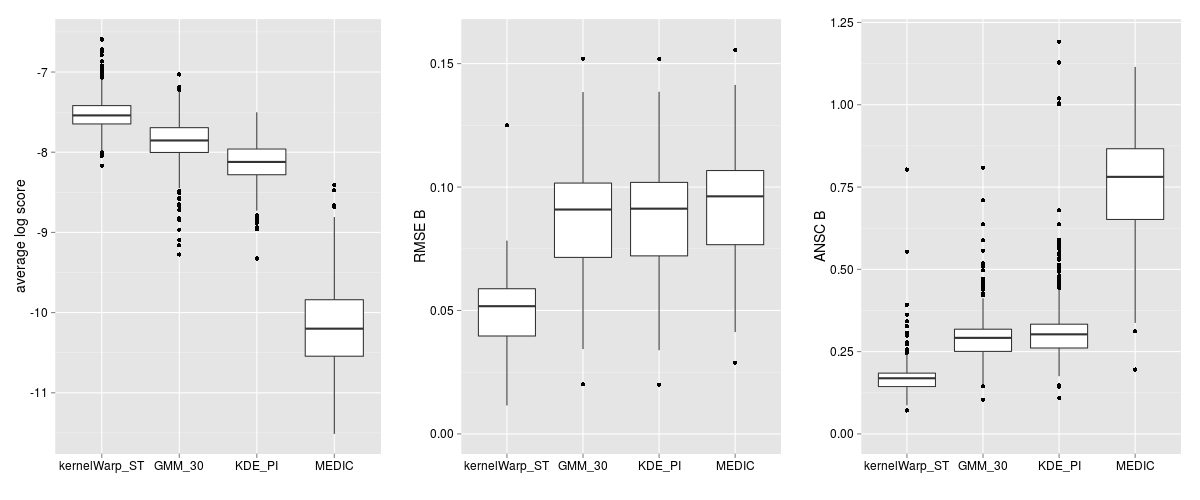}
\caption{Box-plots of predictive accuracies of kernel warping (S-T parameters), GMM (30 comp), KDE (PI bandwidth), and the MEDIC method (an industry practice) over 672 test periods, as measured by average log score (left, less negative is better), $\mbox{RMSE}_B$ (middle, smaller is better), and $\mbox{ANSC}_B$ (right, smaller is better).}
\label{fig:boxplot}
\end{figure}

Additionally, since these cell counts (mostly 0s and 1s) are more appropriately modeled by a discrete distribution such as the Poisson distribution, we also compute the root-mean-square Anscombe residuals \citep{Anscombe:1953, McCullagh:1989}, which specifically adjusts to measure predictive accuracy for Poisson data. Similarly, we consider within all of the rectangular window (ANSC) and within the boundary of Melbourne ($\mbox{ANSC}_B$). Using the same notations as above,
\[
 \mbox{ANSC } (u)=\sqrt{\frac{1}{C}\sum_{c=1}^{C}\left(\frac{(3/2)(y_{u,c}^{2/3}-\hat y_{u,c}^{2/3})}{\hat y_{u,c}^{1/6}}\right)^2},
\]
and $\mbox{ANSC}_B$ is similarly defined. We show in Table \ref{tab:result} the mean predictive accuracies of various methods, averaged across all test time periods $\mathcal{T}_{test}$ (all 1-hour periods in March 2011). A less negative ALS, and smaller RMSE, $\mbox{RMSE}_B$, ANSC, and $\mbox{ANSC}_B$ indicate better predictive accuracy. Both versions of kernel warping have a significant advantage over the comparison methods in all performance measures, especially in $\mbox{RMSE}_B$ and $\mbox{ANSC}_B$. Between the two versions of kernel warping, allowing parameters to be location-specific (in addition to being time-specific) provides additional benefits, even though a large number of time periods choose to use only 1 spatial component. We further show in Figure \ref{fig:boxplot} the box-plots illustrating the variations of some of these metrics across time periods. Kernel warping has not only the best mean performance, but also the smallest variations across time periods. 

\begin{table} [ht]
\hspace*{-0.5cm} 
\begin{tabular}{l c c c c c c} 
\hline
\multicolumn{2}{c}{\textbf{Prediction method}}  & \textbf{ALG} &  \textbf{RMSE} & \textbf{$\mbox{RMSE}_B$} & \textbf{ANSC} & \textbf{$\mbox{ANSC}_B$}  \T\B \\ 
\hline
Kernel warping &  S-T param & $-7.53$ & $0.0500$ & $0.0498$ & $0.176$ & $0.171$ \T \\   
& T param & $-7.56$ & $0.0518$ & $0.0514$ & $0.178$ & $0.172$ \B \\ 
\hline
(a) MEDIC  & & $-10.11$ & $0.0589$ & $0.0996 $ & $0.479$ & $0.810$ \T \B \\  
(b) Unwarped KDE & PI & $-8.14$ & $0.0562$ & $0.0950$ & $0.199$ & $0.334$ \T \\
& SCV & $-8.15$ & $0.0562$ & $0.0950$ &  $0.194$ & $0.325$ \B \\ 
(c) GMM & 15 comp & $-7.96$ & $0.0562$ & $0.0949$  & $0.181$ & $0.304$ \T \\
& 30 comp & $-7.87$ & $0.0561$ & $0.0948$ & $0.191$ & $0.323$ \\
& 50 comp & $-7.93$ & $0.0561$ & $0.0949$ & $0.188$ & $0.316$\B \\ 
\hline
\vspace{0.1cm}
\end{tabular}
\caption{Mean predictive accuracies across all 1-hour periods in March 2011 of the proposed kernel warping and competing methods. Kernel warping outperforms the competing methods.}
\label{tab:result}
\end{table}

\section{Conclusions} \label{sec:concl}

Fine-resolution spatio-temporal ambulance demand predictions are critical to optimal ambulance planning. Typical challenges include data sparsity at the prediction resolution and incorporation of complex urban spatial domains. These challenges are especially prominent in Melbourne. They create a tension; overcoming sparsity requires considerable smoothing, while representing complex spatial features requires fine resolution modeling. Most current industry practices and earlier studies are ill-equipped to address these challenges simultaneously. We propose a kernel warping method that smooths intelligently towards geographical characteristics. We demonstrate that our proposed method predicts ambulance demand in Melbourne more accurately than the state of the art in the practice and research of ambulance demand prediction. 

To predict ambulance demand for any hour, we use a spatio-temporal kernel density estimator on the sparse set of most similar labeled data, but warp these kernels to a larger point cloud drawn from all historical observations regardless of labels. We construct an adjacency graph on this point cloud to approximate Melbourne's spatial boundaries, neighborhoods, and road networks in a data-driven manner. Kernels on labeled data are warped to encourage flow along and penalize flow orthogonal to this graph. 

Kernel warping circumvents the need to define boundaries and boundary conditions, which are often difficult in the practice of modeling point patterns on complex spatial domains. It also captures and exploits finer-grain internal spatial structures other than boundary features, which can be prominent in various heterogeneous environments such as cities, buildings, mountains, and forests. Kernel warping is not limited to density estimation. It can be adapted to model a wide range of functions and surfaces. It can be used to perform a broad set of tasks including prediction, classification, clustering, and visualization. Inferences on uncertainty, if desired, can be obtained by assessing cross-validation variance and warping kernels to different samples of point clouds. 
There is much flexibility in designing the point cloud and its Laplacian. We offer some discussions on these in the context of spatial and spatio-temporal point patterns. We also offer efficient estimation of kernel bandwidth and degree of warping local to time periods and locations via cross-validation. The proposed method is straightforward to implement and easy to experiment with. The tools we have developed can be easily generalized to model a wide range of spatial or spatio-temporal point process on complex spatial domains.

\section*{Acknowledgements} The authors sincerely thank Ambulance Victoria for sharing their data. We also thank the support from a Xerox Faculty Research Award and NSF grant DMS-1455172.

\bibliographystyle{imsart-nameyear}
\bibliography{ZZthesisbib}

\end{document}